\documentclass[a4paper,aps,pra,showpacs,twocolumn]{revtex4-1}

\usepackage{amssymb}
\usepackage{graphicx}
\usepackage{amsmath}
\usepackage{amsbsy}
\usepackage{amsthm}
\usepackage{bbm}
\usepackage{bm}
\usepackage{epsfig}
\usepackage{epstopdf}
\usepackage{dsfont}
\usepackage{hyperref}
\usepackage{soul}
\usepackage{color}

\newcommand{\tr}{\textnormal{Tr}}

\newcommand{\be}{\begin{equation}}
\newcommand{\ee}{\end{equation}}
\newcommand{\beq}{\begin{eqnarray}}
\newcommand{\eeq}{\end{eqnarray}}

\newcommand{\ket}[1]{\ensuremath{| #1 \rangle}}

\begin{document}

\title{Residual entanglement of accelerated fermions is not nonlocal}

\author{Nicolai Friis$^{1}$}
\email{pmxnf@nottingham.ac.uk}
\author{Philipp K\"ohler$^{2}$}
\email{philipp.koehler@gmx.at}
\author{Eduardo Mart$\acute{\i}$n-Mart$\acute{\i}$nez$^{3}$}
\email{emmfis@gmail.com}
\author{Reinhold A. Bertlmann$^{2}$}
\affiliation{$^{1}$School of Mathematical Sciences, University of Nottingham, University Park,
Nottingham NG7 2RD, United Kingdom}
\affiliation{$^{2}$Faculty of Physics, University of Vienna, Boltzmanngasse 5, 1090 Vienna, Austria}
\affiliation{$^{3}$Institute for Quantum Computing, University of Waterloo, 200 Univ. Avenue W, Waterloo ON N2L 3G1, Canada}
%\affiliation{$^{3}$Instituto de F$\acute{\i}$sica Fundamental, CSIC, Serrano 113-B, 28006 Madrid, Spain}
\date{December 2011}
\begin{abstract}
We analyze the operational meaning of the residual entanglement in non-inertial fermionic systems in terms of the achievable violation of the Clauser-Horne-Shimony-Holt (CHSH) inequality. We demonstrate that the quantum correlations of fermions, which were previously found to survive 
in the infinite acceleration limit, 
%[P. Alsing, I. Fuentes-Schuller, R. Mann and T. E. Tessier, Phys. Rev. A \textbf{74}, 032326 (2006)],
cannot be considered to be non-local. The entanglement shared by an inertial and an accelerated observer cannot be utilized for the violation 
of the CHSH inequality in case of high accelerations. Our results are shown to extend beyond the single mode approximation commonly used in the 
literature.
\end{abstract}

%\keywords{}

\maketitle

\section{Introduction}

The extension of quantum information theory to relativistic settings has been a thriving area of research for some time \cite{alsingfuentesschullermanntessier06,ahnleemoonhwang03,terashimaueda03,friisbertlmannhuberhiesmayr10,gingrichadami02peresterno04jordanshajisudarshan07huberfriisgabrielspenglerhiesmayr11,martinmartinezfuentes11,alsingmilburn03,bruschiloukomartinmartinezdraganfuentes10,fuentesschullermann05}. The effects of relativistic motion on entanglement and quantum information protocols have been studied extensively for inertial observers \cite{ahnleemoonhwang03,terashimaueda03,friisbertlmannhuberhiesmayr10,gingrichadami02peresterno04jordanshajisudarshan07huberfriisgabrielspenglerhiesmayr11}, as well as for accelerated systems \cite{alsingfuentesschullermanntessier06,martinmartinezfuentes11,alsingmilburn03,bruschiloukomartinmartinezdraganfuentes10,fuentesschullermann05}. However, while the effects on entanglement have been analyzed in detail for both cases, the implications for tests of non-locality via the violation of Bell inequalities have received treatment only for the case of inertial motion \cite{ahnleemoonhwang03,terashimaueda03,friisbertlmannhuberhiesmayr10}. In the other situation, i.e., if (at least) one of the observers sharing a bipartite entangled state is moving with uniform acceleration, two scenarios naturally arise. The common quantum state can be entangled with respect to bosonic or, on the other hand, fermionic modes. The entanglement degradation of such a two mode state by accelerated motion, commonly attributed to the thermalization due to the Unruh effect \cite{crispinohiguchimatsas08}, has been investigated for bosons \cite{fuentesschullermann05,bruschiloukomartinmartinezdraganfuentes10} and fermions \cite{alsingfuentesschullermanntessier06,bruschiloukomartinmartinezdraganfuentes10,martinmartinezfuentes11} respectively. The distinguishing feature of the fermionic from the bosonic case in these results is found to be a non-zero residual entanglement between the (anti-)fermionic modes in the infinite acceleration limit. The same result has been obtained for the Hawking effect in eternal black holes \cite{martinmartinezgarayleon10}. Moreover, the remaining entanglement cannot be attributed to bound entanglement, since, even in the cases where bound entanglement can feature in principle, the entanglement persists even when measured by the negativity, a measure which only detects distillable entanglement \cite{vidalwerner02}.

It is the aim of this paper to shed light on this issue and assign operational meaning to the residual fermionic entanglement by applying the criterion introduced in Ref.~\cite{horodecki-rpm95} to quantify the maximally possible violation of the Clauser-Horne-Shimony-Holt (CHSH) inequality \cite{chsh69}. We find that the quantum correlations remaining in the infinite acceleration limit cannot be used by the observers to demonstrate quantum non-locality, showing that acceleration effectively degrades correlations even in the fermionic case.

\section{Setting}

\begin{figure}[htp!]
\centering
\includegraphics[width=0.45\textwidth]{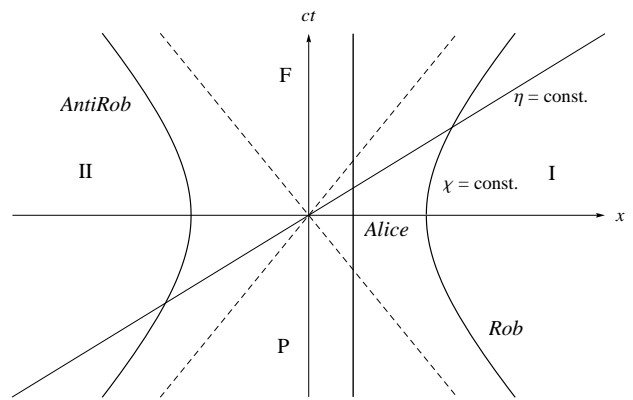}
\caption{The uniformly accelerated observers Rob and \mbox{AntiRob} are confined to the Rindler wedges $\mathrm{I}$ ($|t|\,<\,x$) and $\mathrm{I\!I}$ ($|t|\,<\,-x$), respectively, which are causally disconnected from each other. Their worldlines are hyperbolas, which correspond to lines of constant $\chi = c^2/a$, where $a$ is their proper acceleration and $0\,<\,\chi\,<\,\infty$.}\label{fig:alicerobantirobsetup}
\end{figure}

We employ a scheme of two observers, Alice and (Anti)Rob, one of which, Alice, is inertial (we can assume without loss of generality that Alice is at rest), while the other one, (Anti)Rob, is uniformly accelerated, in complete analogy to the setup used in Ref.~\cite{martinmartinezfuentes11} (see Fig.~\ref{fig:alicerobantirobsetup}).
To describe the point of view of an accelerated observer we introduce the so-called Rindler coordinates $(\eta,\chi)$, which are the proper coordinates of an accelerated observer moving with fixed acceleration $a$. These coordinates relate to Minkowskian coordinates $(t,x)$ by %\footnote{Note that we are not using the conformal Rindler coordinates $t=c\,a^{-1}e^{a\xi/c^2}\sinh\left(\frac{a\tau}{c}\right)$ together with $x=c^{2}a^{-1}e^{a\xi/c^2}\cosh\left(\frac{a\tau}{c}\right)$ (quite common in the literature) but the proper coordinates of an accelerated observer of acceleration $a$ such that the proper lengths and times measured in these units directly correspond to physical distances and time intervals.}
\begin{equation}
ct=\chi \sinh\left(\frac{a\eta}{c}\right),\qquad x=\chi\cosh\left(\frac{a\eta}{c}\right).
\label{eq:change}
\end{equation}
From (\ref{eq:change}) it can be seen, that for constant $\chi$ these coordinates describe hyperbolic trajectories in space-time, which asymptotically approach the light cone.

Observers eternally in uniform acceleration are always restricted to either region $\mathrm{I}$ or $\mathrm{I\!I}$ of the space-time, i.e., the light cone constitutes an effective event horizon for these observers. A quick inspection reveals that the Rindler coordinates defined in (\ref{eq:change}) do not cover the whole Minkowski space-time. Instead these coordinates only cover the right wedge (region $\mathrm{I}$ in Fig.~\ref{fig:alicerobantirobsetup}). To cover the complete Minkowski space-time we need three more sets of coordinates, $ct=-\chi \sinh\left(\frac{a\eta}{c}\right)$, and $x=-\chi\cosh\left(\frac{a\eta}{c}\right)$, for region $\mathrm{I\!I}$, corresponding to an observer accelerating leftward with regard to the Minkowskian origin, as well as $ct=\pm\xi \cosh\left(\frac{a\tau}{c}\right)$, $x=\pm\xi\sinh\left(\frac{a\tau}{c}\right)$ for regions F and P. For both relevant regions ($\mathrm{I}$ and $\mathrm{I\!I}$), the coordinates $(\chi,\eta)$ take values in the whole domain $(-\infty,+\infty)$, thus admitting completely independent procedures of canonical field quantization.

We can now expand the field in terms of a complete set of solutions of the Dirac equation in Minkowski coordinates $(i\gamma^{\mu}\partial_{\mu}+m)\phi=0$, or, instead, in terms of a complete set of solutions of the Dirac equation in Rindler coordinates for regions ($\mathrm{I}$ and $\mathrm{I\!I}$), i.e.,  $[i\gamma^{\mu}(\partial_{\mu}-\Gamma_{\mu})+m]\phi=0$,
where $\gamma^{\mu}$ are the Dirac-Pauli matrices and $\Gamma_{\mu}$ is the spinorial
affine connection.

Hence, the field operator $\phi$ can be expressed as
\begin{small}
\begin{eqnarray}
\phi	&=&	N_\mathrm{M}\sum\limits_{k}
					\left(c_{k,\mathrm{M}}\,u^{+}_{k,\mathrm{M}} + d_{k,\mathrm{M}}^{\dagger}\ u^{-}_{k,\mathrm{M}}\right)
	\label{eq:fermfieldminm7}\\
	&=&	N_{\mathrm{R}}\sum\limits_{j}
			\left(c_{j,\mathrm{I}}\,u^{+}_{j,\mathrm{I}} + d_{j,\mathrm{I}}^{\dagger}\,u^{-}_{j,\mathrm{I}} +
						c_{j,\mathrm{I\!I}}\,u^{+}_{j,\mathrm{I\!I}} + d_{j,\mathrm{I\!I}}^{\dagger}\,u^{-}_{j,\mathrm{I\!I}}\right),
	\nonumber
\end{eqnarray}
\ignorespacesafterend
\end{small}
where $N_\mathrm{M},N_\mathrm{R}$ are normalization constants. The label $u^{\pm}_{k,\mathrm{M}}$ denotes positive and negative energy solutions (particles and antiparticles) with regard to the Killing vector field $\partial_{t}$, whereas $u^{\pm}_{k,\mathrm{I}}$ and $u^{\pm}_{k,\mathrm{I\!I}}$  are the positive and negative frequency solutions of the Dirac equation in Rindler coordinates w.r.t. the corresponding timelike Killing vector field in regions $\mathrm{I}$ and $\mathrm{I\!I}$, respectively. By $c_{j,\Sigma},d_{j,\Sigma}$ with $\Sigma=\mathrm{M},\mathrm{I},\mathrm{I\!I}$ we denote the Minkowski and Rindler particle/antiparticle operators, satisfying the usual anticommutation relations.  The label $k$ is a multilabel including frequency and spin $k=\{\Omega,s\}$, where $s$ is the component of the spin along the quantization axis.

The annihilation operators $c_{k,\mathrm{M}},d_{k,\mathrm{M}} $ define the Minkowski vacuum $\ket{0}_{\mathrm{M}}$ which must satisfy
$c_{k,\mathrm{M}}\ket{0}_\mathrm{M}=d_{k,\mathrm{M}}\ket{0}_\mathrm{M}=0$, $\forall k$. In the same fashion $c_{j,\Sigma},d_{j,\Sigma}$, define the Rindler vacua in regions $\Sigma=\mathrm{I},\mathrm{I\!I}$.

From (\ref{eq:fermfieldminm7}) we extract the transformation  between the Minkowski and Rindler modes
\begin{small}
\begin{equation}
 u^{+}_{j,\mathrm{M}}=\sum\limits_{k}\left[
	\alpha^{\mathrm{I}}_{jk}\,u^{+}_{k,\mathrm{I}} + \beta^{\mathrm{I}}_{jk}\,u^{-}_{k,\mathrm{I}}+
	\alpha^{\mathrm{I\!I}}_{jk}\,u^{+}_{k,\mathrm{I\!I}} + \beta^{\mathrm{I\!I}}_{jk}\, u^{-}_{k,\mathrm{I\!I}}\right].
\end{equation}
\end{small}
The Bogoliubov coefficients that relate both sets of modes are given by the inner product $(u_{k},u_{j})=\int\!d^{3}\!x\,u_{k}^{\dagger}u_{j}$ and are obtained after some elementary but lengthy algebra (see Refs.~\cite{jaureguitorreshacyan91,langlois04,bruschiloukomartinmartinezdraganfuentes10,martinmartinezfuentes11}).

For fixed acceleration, combinations of Minkowski modes can be found, which transform into monochromatic Rindler modes \cite{bruschiloukomartinmartinezdraganfuentes10,martinmartinezfuentes11}. These modes, which share the same vacuum state as the standard monochromatic Minkowski modes, are called Unruh modes, and their associated annihilation operators are
\begin{small}
\begin{subequations}
    \label{eq:Unruhopm7}
	\begin{align}
		C_{k,\mathrm{R}}\,  &\equiv\,\left(\cos r_{k}\,c_{k,\mathrm{I}}-\sin r_{k}\, d^{\dagger}_{k,\mathrm{I\!I}}\right),\\
        C_{k,\mathrm{L}}\,  &\equiv\,\left(\cos r_{k}\,c_{k,\mathrm{I\!I}}-\sin r_{k}\, d^{\dagger}_{k,\mathrm{I}}\right),
	\end{align}
\end{subequations}
\ignorespacesafterend
\end{small}
where $\tan r_{k}=e^{-\pi c \Omega/a}$. We will go beyond the single-mode approximation \cite{alsingmilburn03} and consider a general Unruh mode in the same fashion as in Ref.~\cite{bruschiloukomartinmartinezdraganfuentes10}, i.e., an arbitrary combination of the two independent kinds of Unruh modes
\begin{small}
\begin{equation}
	c_{k,\mathrm{U}}^{\dagger}=
	q_{\mathrm{L}}(C^{\dagger}_{\Omega,\mathrm{L}}\otimes\mathds{1}_{\mathrm{R}})+
	q_{\mathrm{R}}(\mathds{1}_{\mathrm{L}}\otimes C^{\dagger}_{\Omega,\mathrm{R}}).
\label{eq:creat}
\end{equation}
\end{small}
The associated single-particle state is obtained by applying the creation operator (\ref{eq:creat}) to the vacuum. As is commonplace \cite{alsingfuentesschullermanntessier06,bruschiloukomartinmartinezdraganfuentes10,martinmartinezfuentes11}, we will work in the Grassmann scalar case, which is the simplest case that preserves the fundamental Dirac characteristics. With the shorthand notation \begin{small}\mbox{$
\left|\,ijkl\,\right\rangle_{\Omega}\,=
\,\left|\,i_{\,\Omega}\,\right\rangle_{\mathrm{I}}^{+}\,\left|\,j_{\,\Omega}\,\right\rangle_{\mathrm{I}}^{-}
\,\left|\,k_{\,\Omega}\,\right\rangle_{\mathrm{I\!I}}^{+}\,\left|\,l_{\,\Omega}\,\right\rangle_{\mathrm{I\!I}}^{-}$},
\end{small} the Unruh vacuum state of mode $\Omega$ can be expressed as
\begin{small}
\begin{eqnarray}
\left|\,0_{\,\Omega}\,\right\rangle_{\mathrm{U}} &=&
    \cos^{2}r_{\Omega}\left|0000\right\rangle_{\Omega}-
    \sin r_{\Omega}\cos r_{\Omega}\left|0110\right\rangle_{\Omega}\nonumber\\[2mm]
&& +\sin r_{\Omega}\cos r_{\Omega}\left|1001\right\rangle_{\Omega}-
    \sin^{2}r_{\Omega}\left|1111\right\rangle_{\Omega}
\label{eq:mode Omega Unruh vacuum}
\end{eqnarray}
\ignorespacesafterend
\end{small}
(see Ref.~\cite{martinmartinezfuentes11}). Likewise, the particle and antiparticle states of Unruh mode $\Omega$ in the Rindler basis are found to be
\begin{small}
\begin{eqnarray}
\left|\,1_{\,\Omega}\,\right\rangle_{\mathrm{U}}^{+} &=&
    q_{R}\left(\,\cos r_{\Omega}\left|1000\right\rangle_{\Omega}-
    \sin r_{\Omega}\left|1110\right\rangle_{\Omega}\,\right)\nonumber\\[2mm]
&& +\,q_{L}\left(\,\cos r_{\Omega}\left|0010\right\rangle_{\Omega}+
    \sin r_{\Omega}\left|1011\right\rangle_{\Omega}\,\right),\nonumber\\[2mm]
\left|\,1_{\,\Omega}\,\right\rangle_{\mathrm{U}}^{-} &=&
    q_{R}\left(\,\cos r_{\Omega}\left|0100\right\rangle_{\Omega}+
    \sin r_{\Omega}\left|1101\right\rangle_{\Omega}\,\right)\nonumber\\[2mm]
&& +\,q_{L}\left(\,\cos r_{\Omega}\left|0001\right\rangle_{\Omega}-
    \sin r_{\Omega}\left|0111\right\rangle_{\Omega}\,\right).
\label{eq:mode Omega Unruh particle and antiparticle state}
\end{eqnarray}
\end{small}

\section{Entanglement and Nonlocality}

Following the same notation as in Ref.~\cite{martinmartinezfuentes11} we are endowing the fermionic Fock space with a particular tensor product structure. This can be problematic when working with entanglement measures, as shown in Ref.~\cite{monteromartinmartinez11}. However, our procedure is free from any ambiguity because we correctly treat the tensor product structure and anticommutation relations when we compute expectation values.

\begin{figure}[htp!]
\centering
\includegraphics[width=0.45\textwidth]{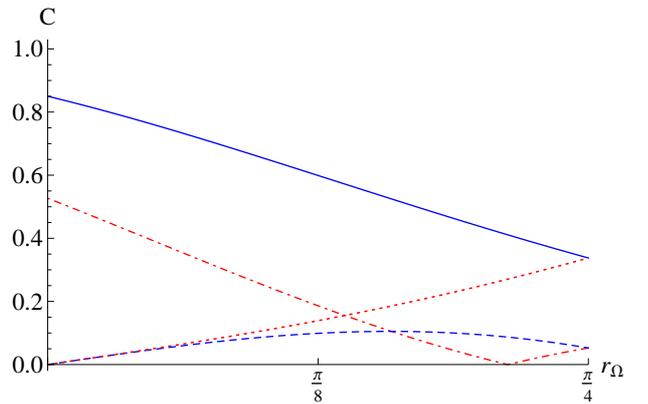}
\caption{(Color online) Concurrence of the reduced states $\rho^{+}_{AR+}$ (blue solid), $\rho^{+}_{AR-}$ (blue dashed), $\rho^{+}_{A\bar{R}+}$ (red dotted-dashed), $\rho^{+}_{A\bar{R}-}$ (red dashed) for $q_{R}=0.85$. Entanglement decreases in the particle sector, which is accompanied by (overall) increases of entanglement in the antiparticle sector. In the infinite acceleration limit, $r_{\Omega}\rightarrow\tfrac{\pi}{4}$, entanglement only vanishes for certain special cases.}\label{fig:concurrenceplot}
\end{figure}

Let us then consider the following initial states
\begin{small}
\begin{equation}
\left|\,\psi_{\pm}\,\right\rangle\,=\, 	
\frac{1}{\sqrt{2}}\,\left(\,
\left|\,0_{\omega}\,\right\rangle_{\mathrm{A}}\,\left|\,0_{\Omega}\,\right\rangle_{\mathrm{U}}\,+\,
\left|\,1_{\omega}\,\right\rangle_{\mathrm{A}}^{\epsilon}\,\left|\,1_{\Omega}\,\right\rangle_{\mathrm{U}}^{\pm}\,\right),
\label{eq:initial state}
\end{equation}
\ignorespacesafterend
\end{small}
where the tensor product structure refers to two distinct inertial observers, Alice and Bob, in Minkowski space-time. We have explicitly assumed here that Bob's mode is a positive ($\psi_{+}$) or negative frequency ($\psi_{-}$) Unruh mode, while Alice's mode (labeled by the subscript ``A") can be either a Minkowski or Unruh mode of positive ($\epsilon = ``+"$) or negative frequency ($\epsilon = ``-"$), respectively. Let us now replace the second observer Bob by the previously discussed accelerated observers Rob or AntiRob. Due to their non-inertial motion, access to Minkowski space is limited. In particular, Rob is causally disconnected from region $\mathrm{I\!I}$, while AntiRob cannot interact with region $\mathrm{I}$, which implies tracing over the unaccessible space-time regions. We additionally make the assumption, that each observer is concerned only with the (positive or negative frequency) mode, originally considered. For each of the initial states we thus obtain four different reduced two-qubit density matrices, computed in Ref.~\cite{martinmartinezfuentes11}. Tracing over region $\mathrm{I\!I}$ and the antiparticle sector (i.e., the negative frequency mode) of the state $\psi_{+}$ in Eq.~(\ref{eq:initial state}), results in
\begin{small}
\begin{equation}
\rho^{+}_{AR+}=\frac{1}{2}\!
	\begin{pmatrix}
\cos^{2}\!r_{\Omega}	&	\hspace*{-0.3cm}0											&	0																	 &	 \hspace*{-0.5cm}q^{*}_{R}\cos r_{\Omega}												\\
		0									&	\hspace*{-0.3cm}\sin^{2}\!r_{\Omega}	&	0																	 &	 \hspace*{-0.5cm}0												\\
		0									&	\hspace*{-0.3cm}0											&	 |q_{L}|^{2}\cos^{2}\!r_{\Omega}		&	 \hspace*{-0.5cm}0												 \\
q_{R}\cos r_{\Omega}	&	\hspace*{-0.3cm}0											&	0																	 &	 \hspace*{-0.5cm}|q_{R}|^{2}+|q_{L}|^{2}\sin^{2}\!r_{\Omega}	\\
	\end{pmatrix}.
\label{eq:psi plus Alice Rob particle sector}
\end{equation}
\end{small}
Likewise, the  reduced state $\rho^{+}_{A\bar{R}+}$ for AntiRob, when he is only able to detect positive frequency modes is obtained by exchanging $q_{R}$ and $q_{L}$ in Eq.~(\ref{eq:psi plus Alice Rob particle sector}). Similarly, the reduced state for Alice and Rob, when Rob's detector registers only antiparticles is given by
\begin{small}
\begin{equation}
\rho^{+}_{AR-}=\frac{1}{2}\!
	\begin{pmatrix}
		\cos^{2}\!r_{\Omega}	&	\hspace*{-0.4cm}0											&	0																							 &	\hspace*{-0.5cm}0												\\
		0											&	\hspace*{-0.4cm}\sin^{2}\!r_{\Omega}	&	 -q^{*}_{L}\sin r_{\Omega}											 &	 \hspace*{-0.5cm}0												 \\
		0											&	\hspace*{-0.4cm}-q_{L}\sin r_{\Omega}	&	 |q_{L}|^{2}+|q_{R}|^{2}\cos^{2}\!r_{\Omega}		&	 \hspace*{-0.5cm}0												 \\
		0											&	\hspace*{-0.4cm}0											 &	0																							 &	 \hspace*{-0.5cm}|q_{R}|^{2}\sin^{2}\!r_{\Omega}	\\
	\end{pmatrix},
\label{eq:psi plus Alice Rob antiparticle sector}
\end{equation}
\ignorespacesafterend
\end{small}
and the corresponding state $\rho^{+}_{A\bar{R}-}$ for AntiRob is obtained from Eq.~(\ref{eq:psi plus Alice Rob antiparticle sector}) by the interchange of $q_{L}$ and $-q_{R}$. Analogously, this procedure can be repeated for the initial state $\psi_{-}$ [see \eqref{eq:initial state}], where the roles of positive and negative frequency modes are exchanged in the reduced density matrices, e.g., $\rho^{-}_{AR+}$ is related to $\rho^{+}_{AR-}$ by a simple sign change of $q_{L}$. For these states a tradeoff in entanglement (in terms of the negativity) between the particle and antiparticle sector has been demonstrated in Ref.~\cite{martinmartinezfuentes11}. We illustrate this behavior by studying another common entanglement measure, the concurrence \cite{wootters95}. The concurrence of a two-qubit density matrix $\rho$ is given by \mbox{$C[\rho]\,=\,\max[0,\sqrt{\lambda_{1}}-\sqrt{\lambda_{2}}-\sqrt{\lambda_{3}}-\sqrt{\lambda_{4}}]$}, where $\lambda_{i}$ are the eigenvalues of the matrix \mbox{$\rho(\sigma_{2}\otimes\sigma_{2})\rho^{*}(\sigma_{2}\otimes\sigma_{2})$} in decreasing order, $\sigma_{2}$ is the second Pauli matrix, and the asterisk denotes complex conjugation. As can be seen in Fig.~\ref{fig:concurrenceplot}, a change in entanglement in the particle sector of Alice and (Anti)Rob for increasing $r_{\Omega}$ is accompanied by an opposite change in the corresponding antiparticle sector, and vice versa.
To better understand the surviving correlations in the the limit $r_{\Omega}\rightarrow\tfrac{\pi}{4}$ we want to infer how the entanglement of the reduced states can be utilized by the observers in an experiment to test Bell inequalities, in particular, the CHSH inequality \cite{chsh69}, which is a suitable inequality for two qubits to test local-realistic theories. Any such theory must satisfy the bound
\begin{equation}
|\left\langle\,\mathcal{B}_{CHSH}\,\right\rangle_{\rho}|\,\leq\,2\, ,
\label{eq:chsh inequality}
\end{equation}
where $\mathcal{B}_{CHSH}=\mathbf{a}\cdot\mathbf{\sigma}\otimes(\mathbf{b}+\mathbf{b^{\prime}})\cdot\mathbf{\sigma}+\mathbf{a^{\prime}}\cdot\mathbf{\sigma}\otimes(\mathbf{b}-\mathbf{b^{\prime}})\cdot\mathbf{\sigma}$, $\mathbf{a},\mathbf{a^{\prime}},\mathbf{b}$, and $\mathbf{b^{\prime}}$ are unit vectors in $\mathbb{R}^{3}$, and $\mathbf{\sigma}$ is the vector of Pauli matrices. It is known that for  some choices of $\mathbf{a},\mathbf{a^{\prime}},\mathbf{b},\mathbf{b^{\prime}}$ inequality (\ref{eq:chsh inequality}) can be violated by certain states $\rho$ up to the value $2\sqrt{2}$. For a general two-qubit state Ref.~\cite{horodecki-rpm95} provides a criterion for the maximally possible violation of (\ref{eq:chsh inequality}). The maximally possible value $\left\langle\,\mathcal{B}_{max}\,\right\rangle_{\rho}$ of the Bell-CHSH expectation value [left-hand side of (\ref{eq:chsh inequality})] for a given state $\rho$ is determined by
\begin{equation}
\left\langle\,\mathcal{B}_{max}\,\right\rangle_{\rho}\,=\,2\,\sqrt{\mu_{1}+\mu_{2}}\, ,
\label{eq:chsh criterion}
\end{equation}
where $\mu_{1},\mu_{2}$ are the two largest eigenvalues of $U(\rho)=T^{T}_{\rho}T_{\rho}$. The matrix $T=(t_{ij})$, where $t_{ij}=\tr[\rho\,\sigma_{i}\otimes\sigma_{j}]$, is the so-called correlation matrix of the generalized Bloch decomposition of the density operator $\rho$. For the reduced state (\ref{eq:psi plus Alice Rob particle sector}) we find
%\begin{small}
%\begin{equation}
%    T_{\rho^{+}_{AR+}}=
%    \begin{pmatrix}
%        \mathrm{Re}(q_{R})\cos r_{\Omega}  &   \!\!\mathrm{Im}(q_{R})\cos r_{\Omega}  &   \!\!0   \\
%        \mathrm{Im}(q_{R})\cos r_{\Omega}  &   \!-\mathrm{Re}(q_{R})\cos r_{\Omega}  &   \!\!0   \\
%        0  &   0  &   \!\!|q_{R}|^{2}\cos^{2}\!r_{\Omega}   \\
%    \end{pmatrix},
%\label{eq:psi plus Alice Rob particle sector correlation matrix}
%\end{equation}
%\end{small}
%from which we immediately get
\begin{small}
\begin{equation}
    U(\rho^{+}_{AR+})=
    \begin{pmatrix}
        |q_{R}|^{2}\cos^{2}\!r_{\Omega}  &   0  &   \!\!0   \\
        0  &   |q_{R}|^{2}\cos^{2}\!r_{\Omega}  &   \!\!0   \\
        0  &   0  &   \!\!|q_{R}|^{4}\cos^{4}\!r_{\Omega}   \\
    \end{pmatrix}.
\label{eq:psi plus Alice Rob particle sector chsh matrix}
\end{equation}
\end{small}
Since $0\leq r_{\Omega}<\tfrac{\pi}{4}$ and $|q_{R}|\leq1$, the first two identical eigenvalues $|q_{R}|^{2}\cos^{2}\!r_{\Omega}$ are always larger than the third eigenvalue. From (\ref{eq:chsh criterion}) we thus obtain
\begin{equation}
\left\langle\,\mathcal{B}_{max}\,\right\rangle_{\rho^{+}_{AR+}}\,=\,2\sqrt{2}\,|q_{R}|\cos r_{\Omega}\, .
\label{eq:chsh criterion for Alice Rob particle}
\end{equation}
Similarly, the maximally possible value of the Bell-CHSH parameter for the antiparticle sector of Alice and Rob is
\begin{equation}
\left\langle\,\mathcal{B}_{max}\,\right\rangle_{\rho^{+}_{AR-}}\,=\,2\sqrt{2}\,|q_{L}|\sin r_{\Omega}\, ,
\label{eq:chsh criterion for Alice Rob antiparticle}
\end{equation}
while the corresponding results for AntiRob are obtained by exchanging $q_{R}$ and $q_{L}$ in (\ref{eq:chsh criterion for Alice Rob particle}) and (\ref{eq:chsh criterion for Alice Rob antiparticle}). %, i.e.
%\begin{eqnarray}
%\left\langle\,\mathcal{B}_{max}\,\right\rangle_{\rho^{+}_{A\bar{R}+}}   &=& 2\sqrt{2}\,|q_{L}|\cos r_{\Omega}\,
%\label{eq:chsh criterion for Alice AntiRob particle}\\
%\left\langle\,\mathcal{B}_{max}\,\right\rangle_{\rho^{+}_{A\bar{R}-}}   &=& 2\sqrt{2}\,|q_{R}|\sin r_{\Omega}\, .
%\label{eq:chsh criterion for Alice AntiRob antiparticle}
%\end{eqnarray}
For the initial state $\psi_{-}$ the results of the particle and antiparticle sectors are simply switched. As can be instantly seen from (\ref{eq:chsh criterion for Alice Rob particle}) and (\ref{eq:chsh criterion for Alice Rob antiparticle}), the bound for local realism [right-hand side of (\ref{eq:chsh inequality})] cannot be surpassed in the infinite acceleration limit $r_{\Omega}\rightarrow\tfrac{\pi}{4}$.
\begin{figure}[htp!]
\centering
\includegraphics[width=0.45\textwidth]{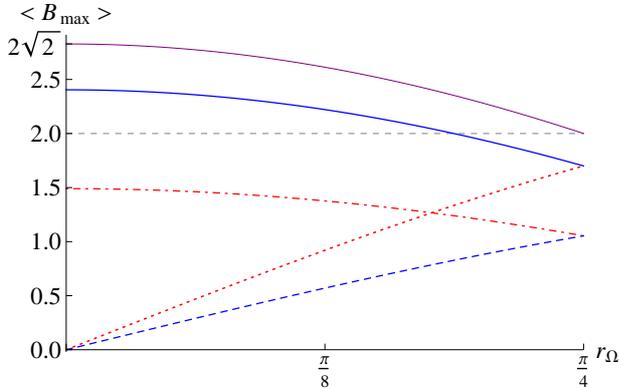}
\caption{(Color online) Maximal Bell-CHSH parameter $\left\langle\,\mathcal{B}_{max}\,\right\rangle$ for the states $\rho^{+}_{AR+}$ (blue solid, second from the top), $\rho^{+}_{AR-}$ (blue dashed), $\rho^{+}_{A\bar{R}+}$ (red dotted-dashed), $\rho^{+}_{A\bar{R}-}$ (red dashed) for $q_{R}=0.85$ and $\rho^{+}_{AR+}$ (purple solid, topmost) for $q_{R}=1$. No non-locality remains in any of the reduced states in the limit $r_{\Omega}\rightarrow\tfrac{\pi}{4}$.}\label{fig:chshplot}
\end{figure}
In fact, for the initial state $\psi_{+}$, the entanglement in the antiparticle sector cannot violate (\ref{eq:chsh inequality}) for any choice of $q_{R},q_{L}$, i.e., $\left\langle\,\mathcal{B}_{max}\,\right\rangle_{\rho^{+}_{AR-}}\leq2$ and $\left\langle\,\mathcal{B}_{max}\,\right\rangle_{\rho^{+}_{A\bar{R}-}}\leq2$, and vice versa for the initial state $\psi_{-}$. Moreover, the CHSH inequality can only ever be violated in the particle sector of either Alice and Rob, or Alice and AntiRob, but never at the same time by both, i.e.,
\begin{small}
\begin{equation}
(\left\langle\,\mathcal{B}_{max}\,\right\rangle_{\rho^{+}_{AR+}}\,+\,\left\langle\,\mathcal{B}_{max}\,\right\rangle_{\rho^{+}_{A\bar{R}+}})/2\,\leq\,2\, .
\label{eq:exclusive nonlocality}
\end{equation}
\end{small}
This can be easily proven, by inserting (\ref{eq:chsh criterion for Alice Rob particle}) and its counterpart for AntiRob into (\ref{eq:exclusive nonlocality}). Because the left-hand side of (\ref{eq:exclusive nonlocality}) is strictly positive, we can consider the square of this expression to get $2\cos^{2}\!r_{\Omega}(1\,+\,2\,|q_{R}|\,|q_{L}|)\,\leq\,2\,(1\,+\,2\,|q_{R}|\,|q_{L}|)$.
Finally, the absolute values can be parameterized as $|q_{R}|=\cos\alpha$ and $|q_{L}|=\sin\alpha$, with $0\leq\alpha\leq\tfrac{\pi}{2}$ and we get $2\,(1\,+\,\sin2\alpha)\,\leq\,4$, which, in turn, implies (\ref{eq:exclusive nonlocality}). Again, the analogous inequality holds for the antiparticle sector if the state $\psi_{-}$ is considered.\\
This exclusiveness of non-locality matches the physical requirements to test a Bell inequality in the accelerated frame. Both observers are required to perform measurements independently of each other and communicate their results. However, in order for (Anti)Rob to be able to receive Alice's measurement results, she has to send these results to (Anti)Rob while still being inside region $\mathrm{I}$ ($\mathrm{I\!I}$). Assuming Alice needs a finite time for this procedure, it is impossible for her to satisfy this requirement for Rob as well as AntiRob simultaneously.\\
Another requirement for such a test of local realism is the ability to use different measurement bases. We assume this to be possible by swapping entanglement to another system, e.g. photons. This can be thought of as a device which produces photons of horizontal polarization when no fermion is incident, while it rotates the photon polarization by $\tfrac{\pi}{2}$, when a (anti)fermion interacts with the apparatus.\\

\section{Conclusions}

We have provided an operational explanation for the surviving fermionic entanglement in infinitely accelerated frames, bridging the gap to the bosonic case, where no entanglement remains in this limit. We have shown that in the fermionic case, the surviving entanglement cannot be used to violate the CHSH inequality, which is the optimal Bell inequality for this situation~\cite{collinsgisinlindenmassarpopescu02}.
Therefore, no quantum information tasks using these correlations can outperform states with appropriate classical correlations. This claim holds if the observers are not allowed to manipulate the final state by local operations or classical communication (LOCC). In particular, maximally entangled, nonlocal states could be distilled from the residual entanglement, if several copies of the state were supplied and the observers could freely communicate. However, this communication is severely limited in the infinite acceleration limit, such that any schemes of recurring local operations (see, e.g., Ref.~\cite{verstraeteverschelde03}) based on the classically transmitted measurement outcomes are excluded.\\
This is especially important not only for the results in the infinite acceleration limit but also if we identify this limit with a black hole situation where an observer is freely falling and another observer is resting arbitrarily close to the event horizon (see Ref.~\cite{martinmartinezgarayleon10}). Alice, when falling into a black hole, cannot communicate on a quantum information level with an observer who is resting near the horizon.\\
We have further found that the choice of Unruh mode (i.e., $q_{R}$ and $q_{L}$) crucially influences which of the accelerated observers, Rob or \mbox{AntiRob}, can in principle test the non-locality of the initial state.\\
Note added. Shortly after submission of this paper Ref.~\cite{smithmann11} appeared, in which similar results, in accordance with this paper, are presented.

\begin{acknowledgements}
We would like to thank \mbox{I. Fuentes} for useful discussions and encouragement and M. Montero for helpful discussions regarding the contents of this paper. N. F. acknowledges support from EPSRC (CAF Grant No. EP/G00496X/2 to I.Fuentes) and the $\chi$-QEN collaboration. E. M.-M. was supported by the CSIC JAE Grant scheme, the Spanish MICINN Project FIS2008-05705/FIS, and the QUITEMAD consortium.
\end{acknowledgements}

\end{document}